\documentclass[aps,prl,twocolumn,groupedaddress,longbibliography]{revtex4-2}

\usepackage{graphicx}
\usepackage{dcolumn}
\usepackage{bm}
\usepackage{amsfonts}
\usepackage{xspace}
\usepackage{color}
\usepackage{epstopdf}
\usepackage{multirow}
\usepackage{amsmath}
\usepackage{float}
\usepackage[colorlinks=true, urlcolor=blue, linkcolor=blue, citecolor=blue]{hyperref}
\usepackage{amsmath}
\usepackage{mathtools}
\usepackage{enumitem}
\usepackage[section]{placeins}
\usepackage{soul}

\begin{document}

\preprint{}

\title{Cavity dark mode mediated by atom array without atomic scattering loss}

\author{Xiaotian Zhang, Zhanhai Yu, Hongrui Zhang, Di Xiang, Hao Zhang}
\email[hzhang@gscaep.ac.cn]{}
\affiliation{Graduate School of China Academy of Engineering Physics, Beijing 100193, China}

\date{\today}

\begin{abstract}
We realize a ring cavity strongly interacting with an atom array with configurable spatial structures. By preparing the atom array with a maximized structure factor, we observe the emergence of a cavity dark mode, where the standing-wave nodes are dynamically locked to the positions of the atoms. The dark mode is decoupled from the atoms, protecting the system from dissipation through atomic scattering, but still mediates strong coupling and enables efficient conversion between two optical modes. Moreover, we impart arbitrary large phase shift on the converted optical fields by translating the atom array. This strongly interacting ring cavity system with single-atom addressability opens ways to quantum optical engineering and the generation of photonic quantum states based on the geometrical structure of atom arrays.

\end{abstract}

\maketitle

Strong atom-light interactions are crucial for advancing quantum information technology \cite{ReisererRMP,ritschColdAtomsCavitygenerated2013}. By using atoms strongly coupled to optical cavities \cite{leonardMonitoringManipulatingHiggs2017a,leonardSupersolidFormationQuantum2017a,vaidyaTunableRangePhotonMediatedAtomic2018,guoOpticalLatticeSound2021} or waveguides \cite{o2013fiber,corzoLargeBraggReflection2016,sorensenCoherentBackscatteringLight2016a,scheuchercirculator,ringrouting,kannanOndemandDirectionalMicrowave2023}, efficient coupling between different optical modes can be achieved. This enables photon switching in various propagation directions \cite{o2013fiber}, atomic Bragg mirrors \cite{corzoLargeBraggReflection2016,sorensenCoherentBackscatteringLight2016a}, all-optical circulators and routers \cite{scheuchercirculator,ringrouting,kannanOndemandDirectionalMicrowave2023}, and large optical phase shifts \cite{beckLargeConditionalSinglephoton2016,StolzRydbergring,VaneeclooRydbergring}. These capabilities are vital for optical quantum engineering \cite{pedrozo-penafielEntanglementOpticalAtomicclock2020,maliaDistributedQuantumSensing2022,JThompsonSpinsqueeze,greveEntanglementenhancedMatterwaveInterferometry2022,liImprovingMetrologyQuantum2023,munizExploringDynamicalPhase2020,Norciasuperradiant,periwalProgrammableInteractionsEmergent2021,cooper2023engineering,youngObservingDynamicalPhases2024} and the advancement of quantum networks \cite{wenlan,hackerPhotonPhotonQuantum2016,niemietzNondestructiveDetectionPhotonic2021,daissQuantumlogicGateDistant2021,niemietzNondestructiveDetectionPhotonic2021}.

To achieve maximum atom-light coupling strength, the most direct approach is to position atoms where the optical field is strongest. In a Fabry-Pérot cavity, this means placing atoms at the standing-wave antinodes \cite{hostenMeasurementNoise1002016,rempesubmicron,davisPhotonMediatedSpinExchangeDynamics2019,yanSuperradiantSubradiantCavity2023,tiancaizhang}. However, at these positions, the strong atom-light coupling is often accompanied by strong spontaneous scattering of light by the atoms into free space, causing photon loss and atomic heating \cite{bochmannLosslessStateDetection2010,zhangCollectiveStateMeasurement2012}. Here, we demonstrate a cavity dark mode that, despite arising from strong atom-light interactions, enables efficient coupling between two optical modes while protecting the atom-cavity system from spontaneous atomic scattering.

The cavity dark mode is analogous to the well-known atomic dark state in Electromagnetically Induced Transparency (EIT) \cite{fleischhauerElectromagneticallyInducedTransparency2005}, where atomic excitation is eliminated through destructive interference of two coupling fields, enabling dissipation-free atomic state transfer as in stimulated Raman adiabatic passage. Photonic dark states have been realized in various systems. For instance, in a coupled-cavity system~\cite{whiteCavityDarkMode2019a,katoObservationDressedStates2019b} where two Fabry-Pérot fiber cavities, each containing an atomic ensemble, are connected by a fiber, a cavity dark mode is formed, separating photonic and atomic excitations into different cavities so that the atoms are not exposed to light. Similarly, optomechanical dark modes \cite{OptomechanicalDarkMode2012,hillCoherentOpticalWavelength2012} have been demonstrated, where a superposition of two cavity modes with different frequencies decouples from the mechanical oscillator, thereby suppressing mechanical dissipation.

\begin{figure}[b]
\includegraphics[width=8.5cm,angle=0]{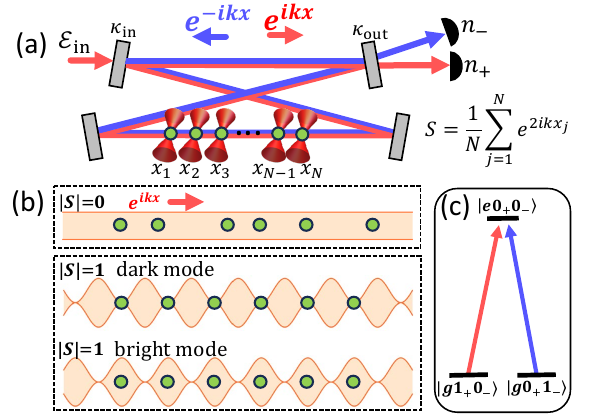}
\caption{\label{fig:1}Experimental schemes. (a) An atom array of $^{87}$Rb atoms with the configurable structure factor $S$ is coupled to an optical ring cavity. We drive the cavity by an input probe beam $\mathcal{E}_{\rm{in}}$ in the forward $+\hat{x}$ direction. Two single-photon counting modules detect the output photons $n_\pm$ of the two cavity traveling-wave $\hat{a}_\pm$ modes $e^{\pm ikx}$. $\kappa_{\rm{in}}$ ($\kappa_{\rm{out}}$) is the coupling rate of the input (output) mirror. (b) When  $S=0$, only the traveling-wave $\hat{a}_+$ mode is excited in the cavity. When $|S|=1$, meaning the atomic separations are integer multiples of half the cavity wavelength $\lambda/2$, the dark/bright mode can be dynamically formed such that all the atoms are located at the nodes/antinodes. (c) The energy level structure illustrates the cavity dark mode. The destructive interference of a coherent superposition of the cavity $\hat{a}_\pm$ modes eliminates atomic excitation.}
\end{figure}

In this work, we realize a cavity dark mode inside a ring cavity that strongly interacts with an atom array \cite{endresatombyatom, Barredoatombyatom} with configurable spatial structures. The ring cavity supports two counter-propagating traveling-wave modes \cite{slamaSuperradiantRayleighScattering2007,ganglColdAtomsHighQ2000,ostermannUnravelingQuantumNature2020,MivehvarDrivenDissipative,klinnerNormalModeSplitting2006a}. Each mode is strongly coupled to the atoms, with the coupling strength determined by the atom array structure factor. When the atom array is configured with a maximized structure factor, with the atomic separations being integer multiples of half the cavity wavelength, a cavity dark mode emerges as a coherent superposition of the two traveling-wave modes, dynamically forming a standing wave with the nodes precisely locked to the atoms. This alignment decouples the atom array from the dark mode, eliminating dissipation from atomic scattering loss. Despite this decoupling, the atom array still mediates strong coupling between the two cavity modes, leading to optical mode conversion. This cavity dark mode is robust against the relative movement between the atom array and the cavity mirrors, as the dark mode can follow the motion of the atom array. By displacing the atom array, we can impart an arbitrarily large phase shift on the converted optical mode.

In our experiment, an in-vacuum optical ring cavity comprising four mirrors has a small waist of $\sim7~\mu$m and finesse of $\mathcal{F}=4.4(1)\times10^4$. Laser-cooled $^{87}$Rb atoms are loaded into a one-dimensional tweezer array at the center of the cavity and further cooled to $5.2~\mu$K by the Raman sideband cooling. The Hamiltonian is given by
\begin{equation} \label{eq:Ham}
\begin{aligned}
&\hat{H}/\hbar=-\delta\left(\hat{a}_+^\dag\hat{a}_++\hat{a}_-^\dag\hat{a}_-\right)
-\Delta\sum_{j=1}^N\hat{\sigma}^+_j\hat{\sigma}^-_j\\
&-\left[ig\sum_{j=1}^N\left(e^{ikx_j}\hat{a}_++e^{-ikx_j}\hat{a}_-\right)\hat{\sigma}^+_j+H.c.\right].
\end{aligned}
\end{equation}
Here $\hat{a}_\pm$ corresponds to the annihilation operator of the linearly-polarized cavity traveling-wave modes $e^{\pm ikx}$, respectively. The coupling between the atom and each of the cavity traveling-wave modes is $g$. We probe the cavity with an input probe beam $\mathcal{E}_{\text{in}}$ in the forward $+\hat{x} $-direction and measure the output photon number $n_{\pm}$ in the $\hat{a}_{\pm}$ modes (Fig.~\ref{fig:1}(a)). The probe and cavity resonance frequencies are tuned near the $5S_{1/2}, F=2\rightarrow5P_{3/2}, F'=3$ transition, with the probe-cavity detuning $\delta$ and the probe-atom detuning $\Delta$. $\hat{\sigma}^\pm_j$ are atomic raising and lowering operators. Atoms are prepared in $|F=2,m_F=2\rangle$ by optical pumping.

Under the weak excitation approximation and adiabatically eliminating the atoms \cite{suppmat}, the equations of motion for the $\hat{a}_{\pm}$ modes are
\begin{equation}\label{eq:eom1}
\begin{aligned}
&\frac{d\hat{a}_+}{dt}=\left(i\delta-\frac{\kappa}{2}+\frac{Ng^2}{i\Delta-\gamma/2}\right)\hat{a}_++\frac{Ng^2S^*}{i\Delta-\gamma/2}\hat{a}_--i\sqrt{\kappa_{\text{in}}}\mathcal{E}_{\text{in}},\\
&\frac{d\hat{a}_-}{dt}=\left(i\delta-\frac{\kappa}{2}+\frac{Ng^2}{i\Delta-\gamma/2}\right)\hat{a}_-+\frac{Ng^2S}{i\Delta-\gamma/2}\hat{a}_+.
\end{aligned}
\end{equation}

Here we characterize the spatial structure of the atom array by the structure factor defined by $S=\sum_{i=1}^N e^{2 i k x_{i}}/N$, which plays a crucial role in mediating the coupling between the two cavity modes $\hat{a}_{\pm}$. We deterministically prepare atom arrays with a variable atom number $N$ and fully controlled positions with a precision of 5 nm by arranging tweezers using the acousto-optic deflectors. Our cavity decay rate $\kappa /(2\pi)= 33.6(8)$ kHz, the atomic decay rate $\gamma /(2\pi)= 6.07$ MHz, and the single-atom cooperativity  $C=4g^2/(\kappa\gamma)=12.5(1)$ for each of the $\hat{a}_\pm$ modes.

\begin{figure}[b]
\includegraphics[width=8.5cm,angle=0]{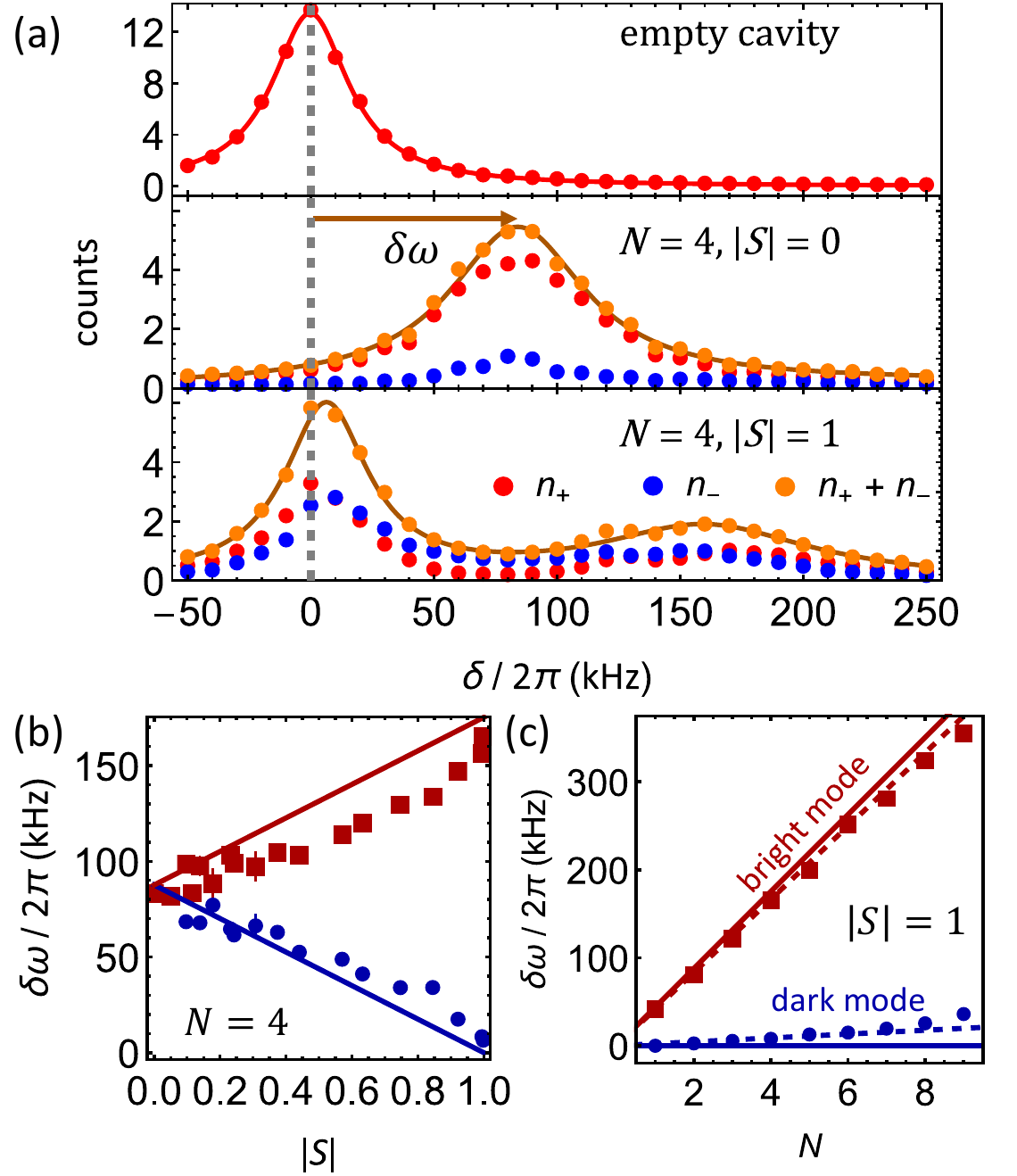}
\caption{\label{fig:2} Cavity dynamical modes tuned by the atom array spatial structures. The cavity forward (red), backward (blue), and total (orange) transmissions are measured for (a) empty cavity and atom arrays with $N=4$ and $|S|=0$, $|S|=1$. The solid lines are Lorentzian fits to total transmission. (b) The cavity shifts of the two modes $\hat{c}_{1,2}$ are measured when varying the atom array structure factor $|S|$. The bright and dark modes emerge when $|S|=1$. The solid lines are the theoretical calculations under ideal conditions with $C=12.5(1)$. (c) At $|S|=1$, the cavity shifts of the bright (square) and dark (circle) modes for atom arrays containing up to 9 atoms. $\Delta/(2 \pi)$ is chosen to be $30\,\text{MHz}$. The solid lines are the theoretical calculations with $C=12.5(1)$ and $|S|=1$. The dashed lines are $|S|=0.9$ considering the finite temperature of atoms. }
\end{figure}

In Eq.(\ref{eq:eom1}), when the structure of the atom array $|S|= 0$, the $\hat{a}_{\pm}$ modes are decoupled and hence are the eigenmodes of the system. Since the cavity is driven in the forward $+\hat{x}$ direction, only the $\hat{a}_{+}$ mode is excited. The cavity forward transmission displays a resonance shifted by $\delta \omega= NC\kappa\gamma\Delta/(4\Delta^2 + \gamma^2) $ relative to the empty cavity and with minimal transmission in the cavity backward direction (Fig.~\ref{fig:2}(a)). Along with the cavity resonance shift, the traveling-wave mode $\hat{a}_{+}$ also suffers the atomic scattering loss, resulting in an increase of the cavity linewidth from $\kappa$ to $\kappa + \delta\kappa=\kappa+ NC\kappa\gamma^2/(4\Delta^2 + \gamma^2)$.

When the structure factor $|S| \neq 0$, the $\hat{a}_{\pm}$ modes are coupled and no longer the eigenmodes. The cavity system is best described in a new basis by performing the transformations
$\hat{c}_1=\frac{1}{\sqrt{2}}\left(\frac{S}{|S|}\hat{a}_++\hat{a}_-\right),
\hat{c}_2=\frac{1}{\sqrt{2}}\left(\frac{S}{|S|}\hat{a}_+-\hat{a}_-\right)$. We obtain a new set of decoupled equations after substituting the transformations into Eq.(\ref{eq:eom1}):
\begin{equation}\label{eq:eom2}
\begin{aligned}
&\frac{d\hat{c}_1}{dt}=\left(i\delta-\frac{\kappa}{2}+\frac{Ng^2\left(1+|S|\right)}{i\Delta-\gamma/2}\right)\hat{c}_1-i\frac{S}{\sqrt{2}|S|}\sqrt{\kappa_{\text{in}}}\mathcal{E}_{\text{in}},\\
&\frac{d\hat{c}_2}{dt}=\left(i\delta-\frac{\kappa}{2}+\frac{Ng^2\left(1-|S|\right)}{i\Delta-\gamma/2}\right)\hat{c}_2-i\frac{S}{\sqrt{2}|S|}\sqrt{\kappa_{\text{in}}}\mathcal{E}_{\text{in}}.
\end{aligned}
\end{equation}
From Eq.~(\ref{eq:eom2}), the $\hat{c}_{1,2}$ modes have the resonance frequencies $\delta \omega_{1,2}= NC (1 \pm|S|) \kappa\gamma\Delta/(4\Delta^2 + \gamma^2)$ and the linewidth broadening $\delta\kappa_{1,2} / \kappa= NC (1 \pm|S|) \gamma^2/(4\Delta^2 + \gamma^2)$. When $ NC |S|$ is large, the cavity exhibits two distinct resonances that are well-separated as shown in Fig.~\ref{fig:2}(a). By fitting the total transmission $n_+ + n_-$ with a sum of two Lorentzian lineshapes, we can determine the photon number $n_{1,2}$ in the $\hat{c}_{1,2}$ modes, respectively. We vary the structure factor $|S|$ and the atom number $N$, and measure the resonance frequencies of the $\hat{c}_{1,2}$ modes for atom arrays containing up to 9 atoms (Fig.~\ref{fig:2}(b-c)).

A special case occurs when the atom array is arranged with $|S|= 1$, meaning the atomic separations are integer multiples of $\lambda/2$. Under this condition, a cavity dark mode emerges, when the resonance of the $\hat{c}_{2}$  mode becomes the same as that of the empty cavity, exhibiting zero cavity shift $\delta \omega=0$ and zero cavity broadening $\delta \kappa/\kappa=0$. The dark mode is analogous to EIT. It is an equal superposition of the $\hat{a}_\pm$ modes, with the phase controlled by the atom array structure factor $S$, such that all the atoms are located at the nodes of the standing wave as shown in Fig. \ref{fig:1}(b). As a result, atoms are shielded from light and introduce no photon loss into free space through atomic scattering, allowing the dark mode to maintain the bare cavity linewidth. Nevertheless, the two counter-propagating traveling waves $\hat{a}_\pm$ are still strongly coupled by the interactions mediated with the atoms. In contrast, the $\hat{c}_{1}$ mode has its antinodes aligned with the atoms, exhibiting a cavity shift and atomic scattering loss twice that of the $|S|= 0$ case, with the single-atom cooperativity of $2C =25.0(2)$, and is referred to as the bright mode.

\begin{figure}[t]
\includegraphics[width=8.5cm,angle=0]{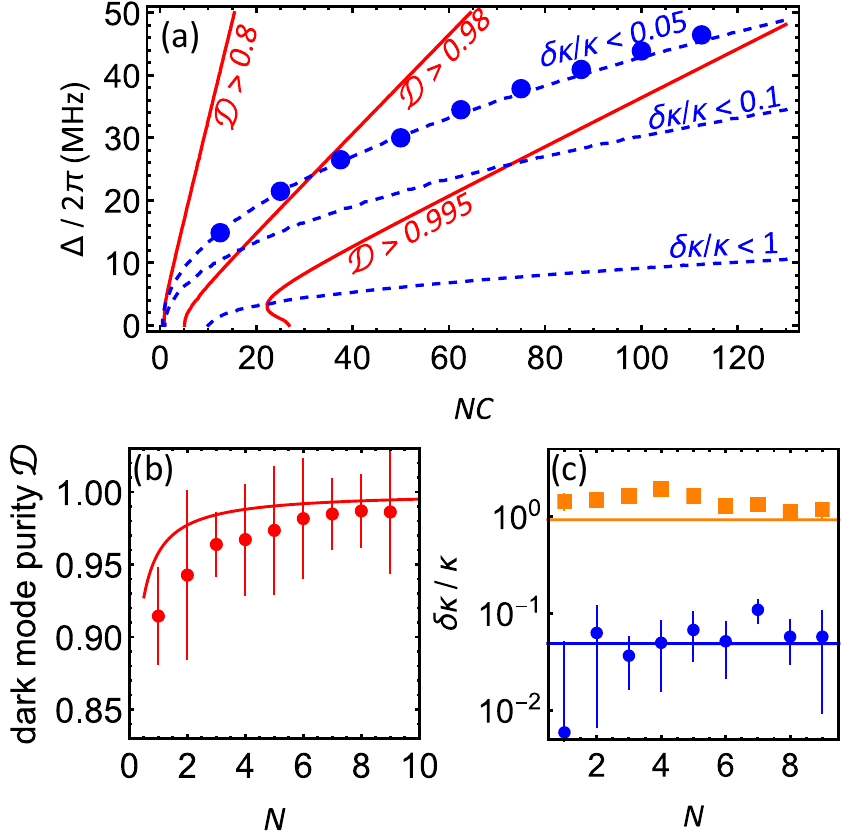}
\caption{\label{fig:3} Characterization of the cavity dark mode by the dark mode purity $\mathcal{D}$ and the atomic scattering broadening $\delta \kappa/\kappa$. (a) Contour plots of $\mathcal{D}$ and $\delta \kappa/\kappa$ with the collective cooperativity $NC$ and the atomic detuning $\Delta$ at $|S|=0.9$. With increasing $NC$, the area to the right of the red contour lines corresponds to larger purity $\mathcal{D}$.
With increasing $\Delta$, the region above the blue contour lines corresponds
to smaller cavity broadening $\delta \kappa/\kappa$.  We vary the atom number $N$ and adjust the corresponding $\Delta$ along the contour line of $\delta \kappa /\kappa =0.05$ indicated by blue circles in (a). The measured dark mode purity $\mathcal{D}$ increases with $N$ (b) while the measured $\delta \kappa/\kappa$ (blue circles) remains constant (c). In comparison, the cavity bright mode (orange squares in (c)) has much larger loss than the dark mode (blue circles). All solid lines are theoretical calculations according to Eqs.(\ref{eq:eom2}).}
\end{figure}
To characterize the quality of the dark mode that we experimentally generate, we measure the dark mode purity $\mathcal{D}$, and the relative linewidth broadening $\delta \kappa /\kappa$ due to atomic scattering loss. The first quantity, the purity $\mathcal{D}$, describes how well the dark mode can be selectively generated while suppressing the excitation of the unwanted lossy bright mode. It is defined as $\mathcal{D}=n_2/(n_2+n_1)$, where $n_{1,2}$ are the photon numbers in the $\hat{c}_{1,2}$ modes, respectively, when the probe beam is tuned to the dark mode resonance frequency $\delta \omega_{2}$. The $n_{1,2}$ values are extracted from the Lorentzian fit of the cavity total transmission in Fig.~\ref{fig:2}(a). In order to suppress the photon number $n_{1}$ from the bright mode leakage, the dark mode needs to be spectrally well-separated from the bright mode. In Fig.~\ref{fig:3}(a) we calculate the purity $\mathcal{D}$ using Eqs.~(\ref{eq:eom2}). At a given atomic detuning $\Delta$, $\mathcal{D}$ increases with the collective cooperativity $N C$. The second quantity, the linewidth broadening $\delta \kappa /\kappa$, describes the level of atomic scattering loss that is achieved in practice. Ideally, the dark mode has $\delta \kappa /\kappa=0$. However, due to the finite spatial spread in traps, atoms do not perfectly remain at the dark mode nodes, resulting in a slight reduction of the structure factor $|S|< 1$ so that $\delta \kappa /\kappa \neq 0$. Through Raman sideband cooling of atoms in our tweezer traps with the radial trapping frequency of $\omega_m=2\pi\times120~\rm{kHz}$, we obtain the atomic spatial spread $\sigma =31~\rm{nm}$, much less than the dark mode standing-wave period of $\lambda/2 =390~\rm{nm}$. This leads to a reduction of the structure factor to  $|S|= 0.9$. We compute $\delta \kappa / \kappa$ in Fig. \ref{fig:3}(a) from Eqs. (\ref{eq:eom2}). At a fixed detuning, $\delta \kappa / \kappa$ increases with $N C$ (Fig. \ref{fig:3}(a)) because the atomic scattering of photons increases. This shows the competing effects of increasing $NC$ at a fixed $\Delta$ when we try to optimize both the dark mode purity $\mathcal{D}$ and the loss $\delta \kappa / \kappa$. However, as long as $NC$ is large enough, we can always adjust $\Delta$ to simultaneously achieve large $\mathcal{D}$ and small $\delta \kappa / \kappa$. Therefore, the collective cooperativity $NC$ is the only fundamental physical quantity that determines the dark mode quality.

In (Fig. \ref{fig:3}(c)), we vary $N$ and correspondingly adjust $\Delta$ to obtain $\delta \kappa / \kappa \sim 0.05$, which is already close to our cavity linewidth measurement resolution limit. Due to our large value of $C=12.5(1)$, we measure $\mathcal{D}>0.98$ with atom arrays containing more than $5$ atoms (Fig. \ref{fig:3}(b)). In comparison, the loss of the bright mode is much larger than the dark mode, with $\delta \kappa / \kappa>1$.

\begin{figure}[h]
\includegraphics[width=6cm,angle=0]{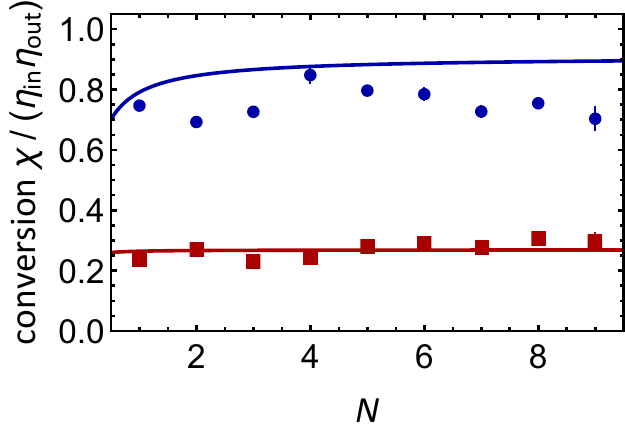}
\caption{\label{fig:4} Conversion $\chi/\eta_{\text{in}}\eta_{\text{out}}$ between two optical modes. The cavity dark mode (blue circles) has a much higher conversion efficiency than the bright mode (red squares). For the dark mode, the conversion efficiency quickly increases to near-unity with the atom number $N>1$, thanks to the large value of $C=12.5(1)$. The conversion is limited by the small linewidth broadening $\delta \kappa/\kappa$, which is due to the finite spatial spread of the atoms. The solid lines are theoretical calculations according to Eqs. (\ref{eq:eom1}).
}
\end{figure}
A high-quality dark mode enables low-loss conversion between the cavity modes. We characterize the overall conversion efficiency as the ratio between the output photon number $n_{-}$ of the backward-traveling mode $\hat{a}_{-}$ and the input photon number $n_{\text{in}}$ in the forward direction, $\chi= n_{-}/n_{\text{in}}$. Using Eqs. (\ref{eq:eom1}), when $NC \gg 1$, the efficiency takes a simple form as $\chi = \eta_{\text{in}}\eta_{\text{out}}/(1+\delta\kappa/\kappa)^2$, where $\eta_{\text{out}}=\kappa_{\rm{out}}/\kappa=0.28$ and $\eta_{\text{in}}=\kappa_{\rm{in}}/\kappa=0.03$ accounting for the output and input photon coupling ratio \cite{suppmat}. In Fig. \ref{fig:4} we show the normalized conversion $\chi/\eta_{\text{in}}\eta_{\text{out}}$, excluding the effects of the finite input and output coupling ratios, as a function of $N$. This normalized $\chi/\eta_{\text{in}}\eta_{\text{out}}$ describes the conversion for the intracavity  $\hat{a}_{-}$ mode. It increases with $N$, reaching $\sim$ 80\% for atom number $N>1$, and is limited by the finite $\delta \kappa / \kappa\sim 0.05$. In contrast, the optical mode conversion operated with the bright mode is much lower due to greater photon loss from atomic scattering. This shows that, in addition to  $\delta \kappa / \kappa$, the conversion efficiency $\chi$ provides a complementary measure of the atomic scattering loss.

The dark mode can also realize arbitrary large optical phase shifts on the converted optical mode. This behavior is formally attributed to the continuous $U(1)$ symmetry inherent in the ring cavity system. Eqs.\eqref{eq:eom1} are invariant under a spatial displacement of the atom array by $x\rightarrow x+X $, accompanied by the cavity phase shifts as $ \hat{a}_{+} \rightarrow \hat{a}_{+}, ~\hat{a}_{-}\rightarrow \hat{a}_{-}e^{2 i k X}$. This means that the intracavity field, given by $\hat{E}(x) \propto \hat{a}_+e^{ikx}+\hat{a}_-e^{-ikx}$, is locked to the atoms and translates together with the displacement of the atoms.

\begin{figure}[h]
\includegraphics[width=8.5cm,angle=0]{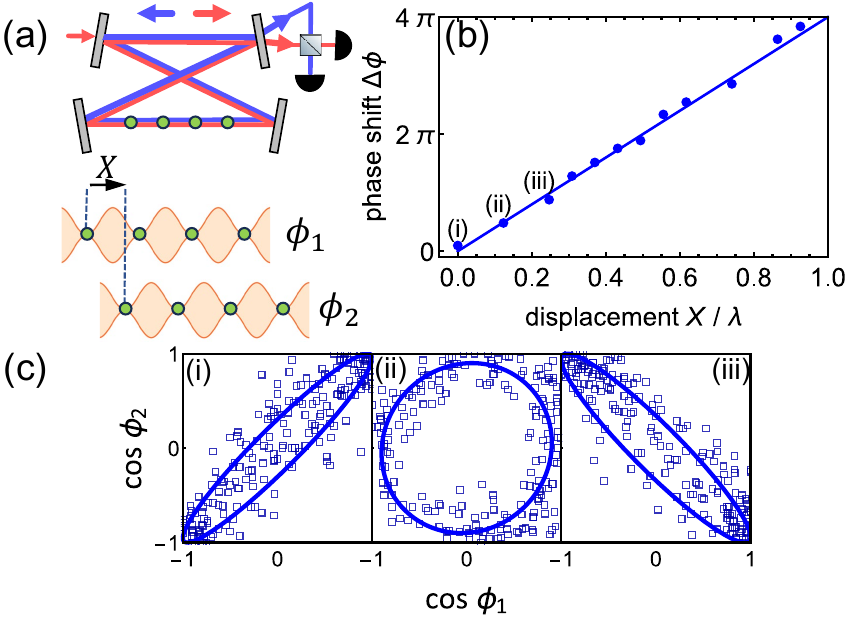}
\caption{\label{fig:5} Large phase shift of the cavity backward field imparted by the atom array. (a) Displace an atom array by distance $X$, and measure the phase shift $\Delta \phi$ of the output of the cavity backward field $\hat{a}_{-}$ through the interference experiment. (b) Arbitrarily large phase shifts can be realized by the atom array displacement. (c) The correlations of $\cos{\phi_2}$ and $\cos{\phi_1}$ before and after the atom array movement that are used to extract $\Delta \phi =\phi_2-\phi_1 $. The solid lines are fit to the data.}.
\end{figure}

To measure the phase $\phi$ of the converted output field of the $\hat{a}_{-}$ mode, we perform an interference experiment by directing the cavity forward and backward transmitted fields to a beamsplitter and measuring the output photon numbers (Fig.~\ref{fig:5}(a)). Displacing the atom array of 4 atoms by a distance $X$ from any initial position, we plot the correlations of $\cos{\phi_1}$ and $\cos{\phi_2}$ in Fig.~\ref{fig:5}(c), where $\phi_1$ and $\phi_2$ are the phases of the cavity output field before and after displacing the atom array, respectively. We obtain the phase change $\Delta \phi = \phi_2 - \phi_1$ from the correlations and confirm that arbitrary large optical phase shift can be achieved as $\Delta\phi = 2 kX$ (Fig.~\ref{fig:5}(b)).

There are several directions for future research with this system. First, while the overall conversion efficiency is currently limited by the coupling ratio of the input and output mirrors, near-unity conversion can be realized by using a one-sided cavity with the same input and output mirror of dominant coupling \cite{liTunableAtomcavityInteractions2024} such that $\eta_{\text{in}}=\eta_{\text{out}}=1$. This will enable efficient generation and routing of directional photons, essential for applications in quantum networks. Second, optical cavities can be used for mid-circuit readout of atomic states \cite{deistMidcircuitCavityMeasurement2022}. However, changes in the atomic state and atom loss limit the readout fidelity. The cavity dark mode, resulting from strong interactions with atoms, carries information of the atomic state while eliminating atomic excitation, offers a new method to improve readout fidelity. Third, the ability of atom arrays to impart arbitrary large phase shifts suggests promising future directions. We plan to utilize this capability to store single photons in the array and realize photon phase gates \cite{VaneeclooRydbergring, StolzRydbergring} with arbitrarily large phases. Lastly, since the ring cavity offers two counter-propagating cavity modes and multiple photon output ports, coupling photons to atom arrays with arbitrary structures will enable the generation of entangled states of many photons \cite{thomasEfficientGenerationEntangled2022} in different spatial directions. This integrated system of an optical ring cavity with atom arrays introduces new possibilities for photon-photon interactions and all-optical quantum information processing based on the tuning geometrical structures of atom arrays.

This work was supported by the National Key Research and Development Program of China (Grant No. 2022YFA1405302) and the National Natural Science Foundation of China (Grants No. 12088101 and No. U2330401).

\bibliography{mainbib}
\clearpage
\onecolumngrid

\setcounter{equation}{0}
\setcounter{section}{0}
\setcounter{figure}{0}
\setcounter{table}{0}
\setcounter{page}{1}

\renewcommand{\theequation}{S\arabic{equation}}
\renewcommand{\thesection}{ \Roman{section}}

\renewcommand{\thefigure}{S\arabic{figure}}
\renewcommand{\thetable}{\arabic{table}}
\renewcommand{\tablename}{Supplementary Table}

\renewcommand{\bibnumfmt}[1]{[S#1]}
\renewcommand{\citenumfont}[1]{S#1}

\section{experiment sequence}
The experiment sequence starts with loading $^{87}$Rb atoms into the tweezer array from a magneto-optical trap (MOT). The tweezer array is generated by focusing an 850 nm laser into Gaussian beams with the waist of $\sim0.9\mu$m through an NA = 0.5 aspherical lens inside the vacuum chamber. The typical trap depth in our experiment is 1.7 mK. After the polarization gradient cooling, the atoms are cooled to $\sim$30 $\mu$K. We then arrange atoms from the randomly loaded array to form specific spatial structures by moving traps with the Acousto-Optic Deflectors (AOD). With Raman sideband cooling, the radial temperature of the atoms are further reduced to 5.2 $\mu$K. Next, a probe beam stabilized to the cavity resonance frequency drives the cavity. Atoms are prepared to $5S_{1/2}, |F=2,m_F=2\rangle$ by optical pumping. Cooling, optical pumping, and cavity probing are repeated 10 times. Finally, the tweezers are turned off to release all the atoms, and the cavity probing is repeated 10 times to detect the empty cavity signal. Throughout the whole process, three fluorescence images are taken to identify the atoms after the initial loading, after arranging them into specific array structures, and to ensure no atoms are lost after the cavity experiment.
\begin{figure}[htbp]
    \centering
    \includegraphics[width=14cm]{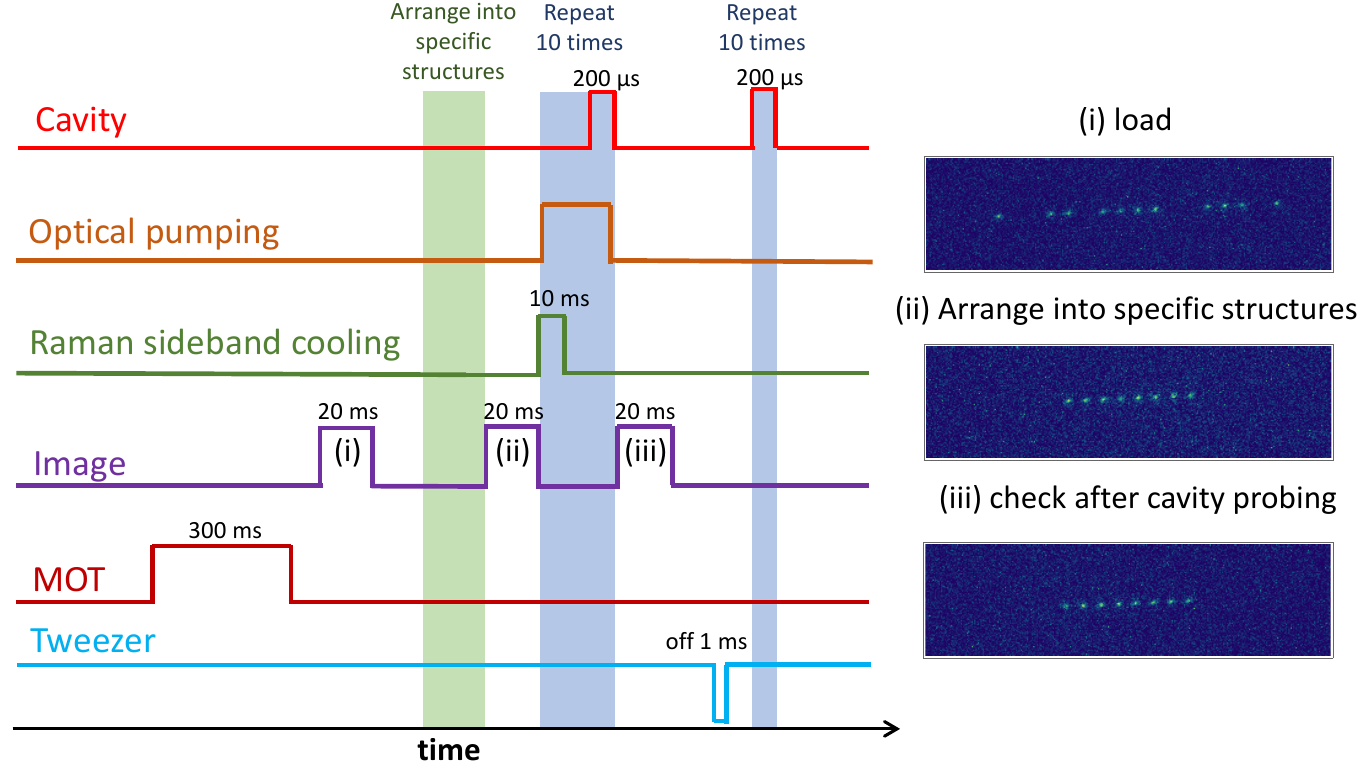}
    \caption{\label{fig:s1}Experimental sequence.}
\end{figure}
\section{Ring cavity eigenmodes and measurement of the cooperativity $C$}
A ring cavity supports two counter-propagating traveling-wave modes $e^{\pm ikx}$ with the corresponding annihilation operators $\hat{a}_+$ and $\hat{a}_-$. When coupling with an atom array, the Hamiltonian is given by
\begin{equation} \label{eqs:Ham}
\begin{aligned}
\hat{H}/\hbar=-\delta\left(\hat{a}_+^\dag\hat{a}_++\hat{a}_-^\dag\hat{a}_-\right)
-\Delta\sum_{j=1}^N\hat{\sigma}^+_j\hat{\sigma}^-_j+\left[-ig\sum_{j=1}^N\left(e^{ikx_j}\hat{a}_++e^{-ikx_j}\hat{a}_-\right)\hat{\sigma}^+_j+H.c.\right]
+\left(\sqrt{\kappa_{\text{in}}}\mathcal{E}_{\text{in}}\hat{a}_+^\dag+H.c.\right).
\end{aligned}
\end{equation}
Where $\delta$ is the probe-cavity detuning and $\Delta$ is the probe-atom detuning. The cavity-atom coupling strength is $g$. The last term $\mathcal{E}_{\text{in}}$ is the input field amplitude from the forward  $+\hat{x} $ direction, and $\kappa_{\text{in}}$ is the coupling rate of the input mirror. Under the approximation that the atoms are weakly excited ($\hat{\sigma}_z\approx-1$) and including the cavity total decay rate $\kappa$ and the atomic spontaneous emission $\gamma$, the equations of motion are given by
\begin{equation}\label{eqs:eom1}
    \begin{aligned}
        &\frac{d\hat{a}_+}{dt}=\left(i\delta-\frac{\kappa}{2}\right)\hat{a}_++g\sum_{j=1}^Ne^{-ikx_j}\hat{\sigma}^-_j-i\sqrt{\kappa_{\text{in}}}\mathcal{E}_{\text{in}},\\
        &\frac{d\hat{a}_-}{dt}=\left(i\delta-\frac{\kappa}{2}\right)\hat{a}_-+g\sum_{j=1}^Ne^{ikx_j}\hat{\sigma}^-_j,\\
        &\frac{d\hat{\sigma}^-_j}{dt}=\left(i\Delta-\frac{\gamma}{2}\right)\hat{\sigma}^-_j-g\left(e^{ikx_j}\hat{a}_++e^{-ikx_j}\hat{a}_-\right).
    \end{aligned}
\end{equation}
In the dispersive regime the atomic internal states $\hat{\sigma}^-_j$ can be adiabatically eliminated according to:
\begin{equation}\label{eqs:sigmass}
    \begin{aligned}
        \langle\hat{\sigma}^-_j\rangle=\frac{g\left(e^{ikx_j}\langle\hat{a}_+\rangle+e^{-ikx_j}\langle\hat{a}_-\rangle\right)}{i\Delta-\gamma/2}.
    \end{aligned}
\end{equation}
Insert Eq.(\ref{eqs:sigmass}) into Eq.(\ref{eqs:eom1}), the cavity modes $\hat{a}_+$ and $\hat{a}_-$ are coupled to each other described by

\begin{equation}\label{eqs:eom2}
    \begin{aligned}
        &\frac{d\hat{a}_+}{dt}=\left(i\delta-\frac{\kappa}{2}+\frac{Ng^2}{i\Delta-\gamma/2}\right)\hat{a}_++\frac{Ng^2S^*}{i\Delta-\gamma/2}\hat{a}_--i\sqrt{\kappa_{\text{in}}}\mathcal{E}_{\text{in}},\\
        &\frac{d\hat{a}_-}{dt}=\left(i\delta-\frac{\kappa}{2}+\frac{Ng^2}{i\Delta-\gamma/2}\right)\hat{a}_-+\frac{Ng^2S}{i\Delta-\gamma/2}\hat{a}_+.
    \end{aligned}
\end{equation}
where $Ng^2/\left(i\Delta-\gamma/2\right)$ describes the cavity shift and broadening for each of $\hat{a}_+$ and $\hat{a}_-$, and $Ng^2S^*/\left(i\Delta-\gamma/2\right)$ ($Ng^2S/\left(i\Delta-\gamma/2\right)$) represents the backscattering from $\hat{a}_-$ ($\hat{a}_+$) to $\hat{a}_+$ ($\hat{a}_-$).

The cavity modes can be made to decouple from each other when we perform the transformations
\begin{equation}\label{eqs:transform}
\begin{aligned}
\hat{c}_1&=\frac{1}{\sqrt{2}}\left(\frac{S}{|S|}\hat{a}_++\hat{a}_-\right),\\
\hat{c}_2&=\frac{1}{\sqrt{2}}\left(\frac{S}{|S|}\hat{a}_+-\hat{a}_-\right).
\end{aligned}
\end{equation}
We obtain a new set of decoupled equations after substituting Eq.(\ref{eqs:transform}) into Eq.(\ref{eqs:eom2}):
\begin{equation}\label{eqs:eom3}
\begin{aligned}
&\frac{d\hat{c}_1}{dt}=\left(i\delta-\frac{\kappa}{2}+\frac{Ng^2\left(1+|S|\right)}{i\Delta-\gamma/2}\right)\hat{c}_1-i\frac{S}{\sqrt{2}|S|}\sqrt{\kappa_{\text{in}}}\mathcal{E}_{\text{in}},\\
&\frac{d\hat{c}_2}{dt}=\left(i\delta-\frac{\kappa}{2}+\frac{Ng^2\left(1-|S|\right)}{i\Delta-\gamma/2}\right)\hat{c}_2-i\frac{S}{\sqrt{2}|S|}\sqrt{\kappa_{\text{in}}}\mathcal{E}_{\text{in}},
\end{aligned}
\end{equation}
where $\hat{c}_1,\hat{c}_2$ are the eigenmodes of the cavity dressed by atoms. The first terms in the right-hand side of Eq.(\ref{eqs:eom3}) indicate that the shifts of the $\hat{c}_1$ and the $\hat{c}_2$ modes are
\begin{equation}\label{eqs:shifts}
\begin{aligned}
&\delta \omega_{1}= \frac{NC\kappa\gamma\Delta}{4\Delta^2 + \gamma^2}(1+|S|),\\
&\delta \omega_{2}= \frac{NC\kappa\gamma\Delta}{4\Delta^2 + \gamma^2}(1-|S|),\end{aligned}
\end{equation}
where we have used the definition of the single-atom cooperativity $C$ for each of the traveling-wave modes $\hat{a}_\pm$, $C = 4g^2/(\kappa \gamma)$. Eq.(\ref{eqs:shifts}) is consistent with the experimental results in Fig.2(b) of the main text, which demonstrates the linear dependence of the cavity shift $\delta\omega_{1,2}$ on $|S|$ for both the $\hat{c}_1$ and the $\hat{c}_2$ modes.

\begin{figure}[h]
    \centering
    \includegraphics[width=16cm]{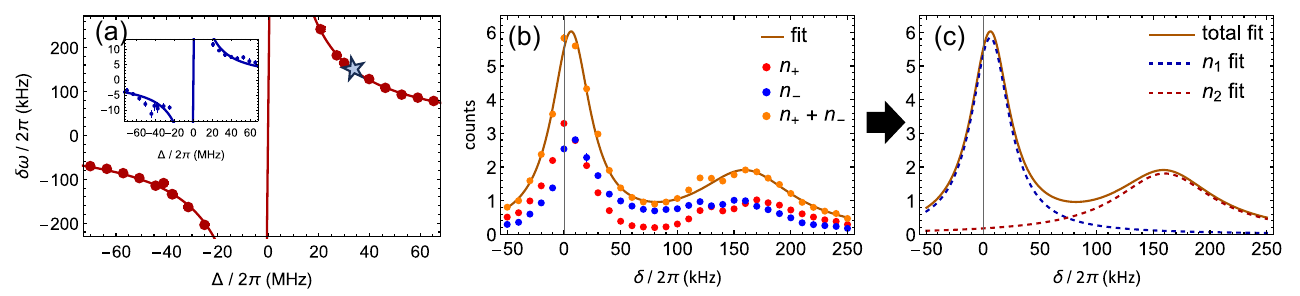}
    \caption{\label{fig:s2} The single-atom cooperativity $C$
is measured from the cavity resonance shift. Drive the cavity coupled to an atom array of $4$ atoms arranged to the maximal structure and measure the transmission spectrum. (a) The cavity shifts of bright (red) and dark (blue, in the inset) modes are measured by varying the probe-atom detuning $\Delta$.  Solid lines are fit to data with the single-atom cooperativity
$C = 12.5(1)$ and $|S|=0.9$. (b) Cavity spectra at $\Delta/2\pi=30$ MHz (indicated by the star in (a)). We measure the transmitted photon numbers $n_{\pm}$ of the $\hat{a}_\pm$ modes and fit the total transmission $n_+ + n_-$ with a double Lorentzian lineshape to determine the shifts of $\hat{c}_{1,2}$ modes. (c) The resonance spectra of the $\hat{c}_{1,2}$ modes (dashed lines) are extracted from the fitted result in (b).}
\end{figure}

To experimentally determine the cooperativity $C$, we measure the dispersive dependence of cavity shifts on the probe-atom detuning $\Delta$ (Fig. \ref{fig:s2}). We position an array with $4$ atoms with the maximal structure factor at the center of the cavity and measure the transmission of the $\hat{a}_\pm$ modes. By fitting the total transmission, $n_++n_-$, with a double Lorentzian lineshape (Fig. \ref{fig:s2}(c)), the resonance
frequencies $\delta\omega_{1,2}$ of bright and dark modes are extracted (Fig. \ref{fig:s2}(b)). Using the structure $|S|=0.9$ due to the finite spatial extent of atoms at the temperature $5.2~\mu\rm{K}$, we obtain the cooperativity as $C=12.5(1)$.

\section{Measurement of cavity parameters}
\begin{figure}[h]
    \centering
    \includegraphics[width=12cm]{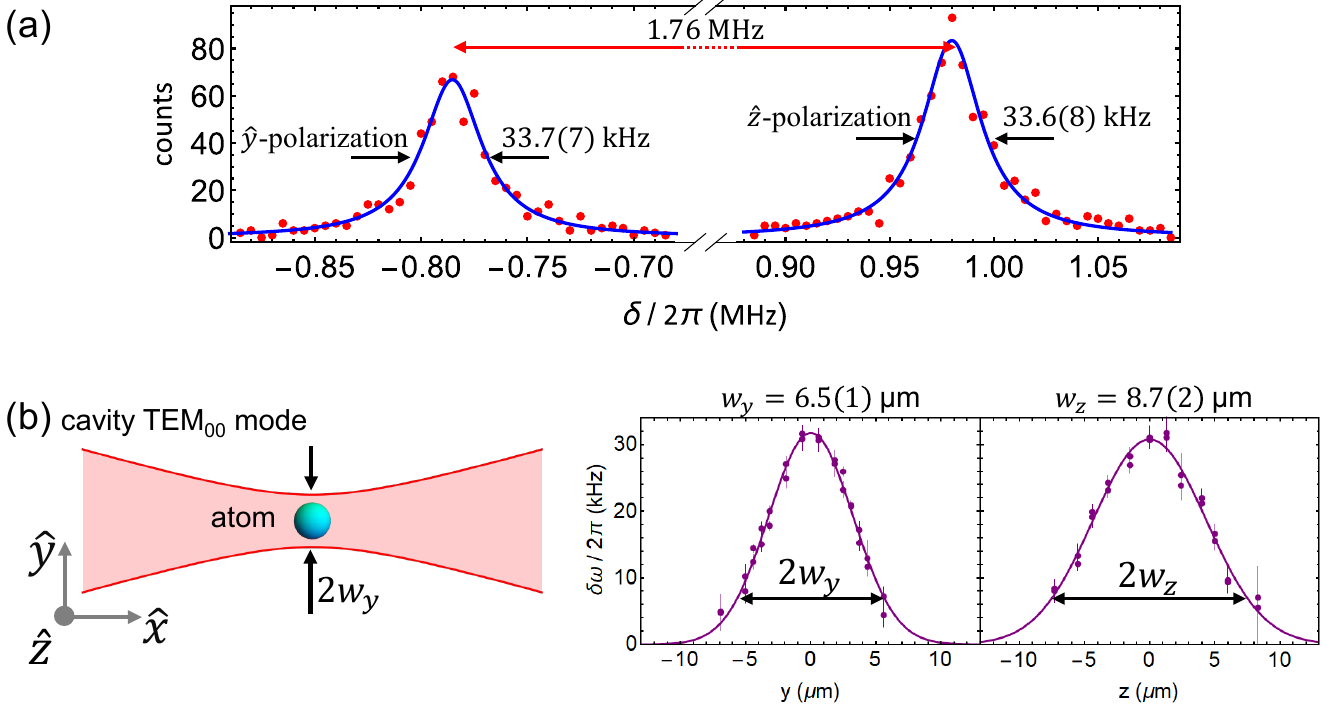}
    \caption{\label{fig:s3} Measurements of the cavity linewidth and the waists. (a) The ring cavity eigenmodes have two orthogonal linear polarization modes due to the non-zero incidence angles, with the birefringence splitting of $\Delta\omega_{\text{bf}}/(2\pi)= 1.76\, \text{MHz}$. Both cavity modes have similar linewidths. The resonance peak heights are not exactly the same because the cavity input polarization in this scan is not perfectly balanced between the two cavity polarization modes. (b) The bright mode frequency shift $\delta\omega$ is measured as the atom is scanned along the $\hat{y}$ and $\hat{z}$ directions. This measurement allows us to determine the waists of the cavity TEM$_{00}$ mode.}
\end{figure}

The ring cavity comprises four mirrors arranged in a bow-tie configuration. Due to the non-zero angle-of-incidence (AOI) of $7^{\circ}$  at each mirror, our ring cavity has a birefringent mode splitting of $\Delta\omega_{\text{bf}} /(2\pi) =1.76$ MHz at 780 nm. The polarization of the two modes is nearly linear, aligned along the horizontal $\hat{y}$ and the vertical $\hat{z}$ directions, respectively. The decay rates are  $\kappa/( 2\pi)= 33.7(7)$ and 33.6(8) kHz for the two polarization modes, respectively, obtained from the measured cavity transmission Lorentzian lineshapes with the probe light stabilized with respect to the cavity (Fig. ~\ref{fig:s3}(a)). This results in similar cooperativity $C$ values for both polarization modes. In all measurements in the main text, we drive the $\hat{z}$-linearly polarized cavity mode. Because $\delta\ll\Delta\omega_{\text{bf}}$,  only one cavity polarization mode is excited.

The cavity free spectral range $\nu_{\text{FSR}}$ = 1472.091(2) MHz is measured by sending both a probe light carrier and a sideband generated by an electro-optic modulator (EOM) and making them simultaneously on resonance with the cavity. Through the measurements of the cavity linewidth $\kappa$ and the free spectral range $\nu_\text{FSR}$, we obtain the cavity finesse $\mathcal{F}=4.4(1)\times10^4$ and the cavity length $L=203.6508(3)$ mm.  The total loss and transmission of all four cavity mirrors, $\mathcal{L}_{\text{total}} + \mathcal{T}_{\text{total}} = 144$ ppm, is determined from the finesse $\mathcal{F}$. One of the cavity mirrors has a measured transmission of $\mathcal{T}=40$ ppm, while all the other three mirrors have nearly identical transmissions of $\sim5$ ppm. All cavity transmission signals are collected from the mirror with the maximum transmission of $\mathcal{T}=40$ ppm, in order to get the maximum signal.

We directly measure the waists of the cavity TEM$_{00}$ mode by using a single atom as a moving probe (Fig. \ref{fig:s3}(b)). The atom is scanned along the $\hat{y}$ and $\hat{z}$ directions by moving the aspherical lens mounted on ultra-high vacuum-compatible translation stages. By measuring bright mode frequency shifts with different atom positions, we can precisely measure the cavity field profile and determine the mode waists of $w_y = 6.5(1) \,\mu\text{m}$ and $w_z = 8.7(2) \,\mu\text{m}$. The cooperativity can be determined from the measured values of the waists and the finesse $\mathcal{F}$:
\begin{equation}
    \begin{aligned}
        C_0=\frac{6\mathcal{F}}{k^2\pi w_yw_z},
    \end{aligned}
\end{equation}
which yields a value of $22.9(8)$. Note this value of $C_0$ corresponds to the maximal coupling by driving the closed transition $5S_{1/2}|F=2,m_F=2\rangle\to5P_{3/2}|F'=3,m_{F'}=3\rangle$ with the circularly-polarized light. Because the ring cavity only supports the linear polarizations, the actual cooperativity in our experiment is reduced by a factor of $\frac{1}{2}$ so $C=\frac{1}{2}C_0=11.5(4)$, which is consistent with the value of $C=12.5(1)$ obtained from the cavity resonance shift measurements in Fig.\ref{fig:s2}(a).
\section{Atom position calibration}
\begin{figure}[h]
    \centering
    \includegraphics[width=12cm]{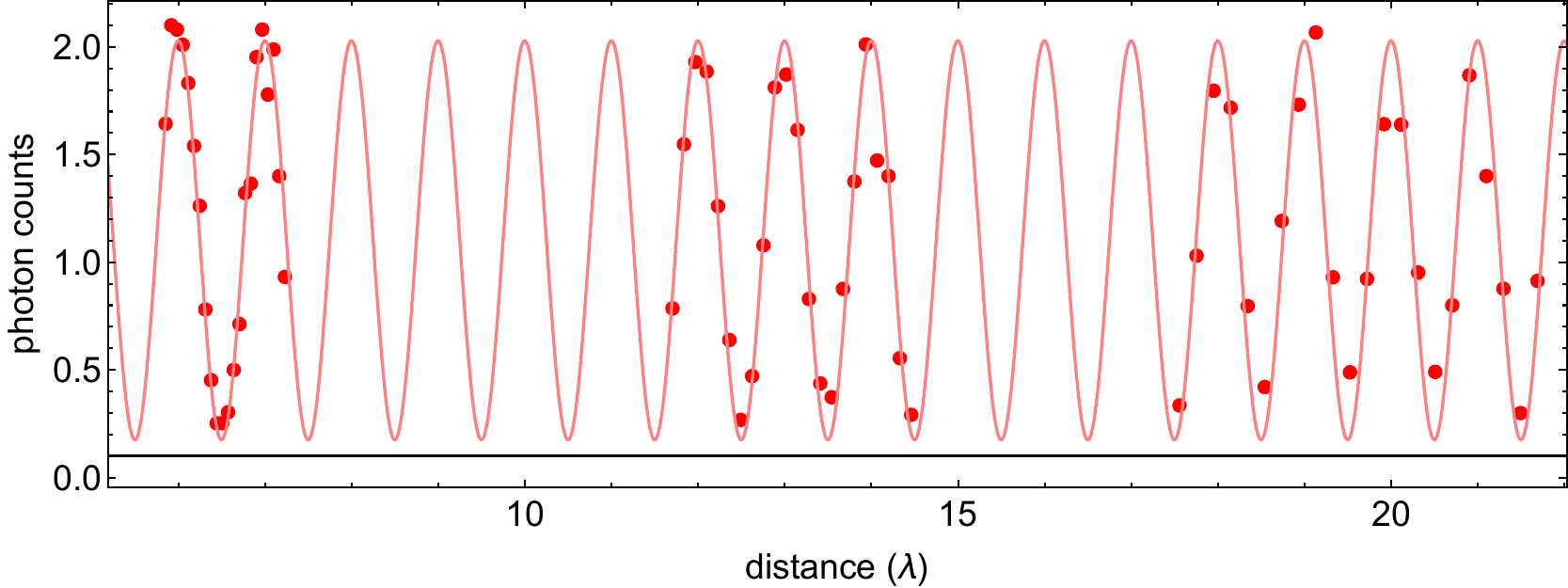}
    \caption{\label{fig:s4}The interference fringes of two atoms are observed by carefully adjusting their separations. The sinusoidal variation of the total photon counts on the two detectors demonstrates the capability to precisely position atoms within the cavity. The black line is the total dark counts of the two detectors. }
\end{figure}

To precisely control the atomic positions and prepare atom arrays with arbitray structure, we control the position of each atom with the precision of 5 nm achieved by adjusting the frequency of the RF tones supplied to the acousto-optic deflector (AOD). Our tweezer array projection optics are designed such that the 1-kHz RF frequency difference on the AOD corresponds to a spatial distance of 5 nm in the tweezer positions. This precision is validated experimentally by measuring the interference of cavity emission from two atoms. We collect the cavity transmission of the $\hat{a}_\pm$ modes when driving two atoms at different separations at a large probe-atom detuning $\Delta$. The total transmitted photon counts show sinusoidal modulation as a function of the atomic distance up to $22\lambda$ (Fig.~\ref{fig:s4}). The observation of maximum and minimum photon counts at even- and odd-multiples of $\lambda/2$ is a clear signature of interference, establishing the atomic distance calibrations for the observation of bright and dark modes in the main text.


\section{temperature effect on the structure factor}
The structure factor $S=1$ when the atomic spacings are perfectly integer multiples of $\lambda/2$. However, the finite spatial extent of atoms reduces the absolute value of the structure factor $|S|$ from unity. The position of each atom follows the Gauss distribution with the probability $p(x)\propto e^{-x^2/(2\sigma^2)}$, where $\sigma$ is the standard deviation of the position distribution and is given by
\begin{equation}
    \begin{aligned}
        \sigma=\sqrt{\frac{\hbar}{2m\omega_m}\coth(\frac{\hbar\omega_m}{2k_BT})},
    \end{aligned}
\end{equation}
where the trap frequency $\omega_m=2\pi\times120~\rm{kHz}$.

\begin{figure}[ht]
\includegraphics[width=8cm,angle=0]{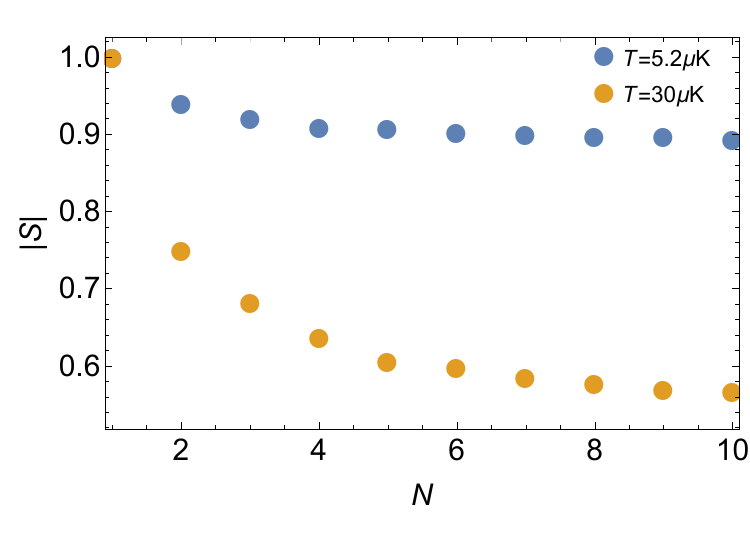}
\caption{\label{fig:s5} Reduction of $|S|$ due to the finite atomic spatial spread at two different temperatures. $|S|$ is reduced from unity to a constant value when the atom number $N$ is large. }
\end{figure}

We cool the atoms close to the ground vibrational state of the tweezer traps using the Raman sideband cooling, with the temperature of $5.2~\mu\rm{K}$ and the mean vibrationanl quantum number $n = 0.4$. At this low temperature, the atomic spatial spread is dominated by the quantum fluctuation and  $\sigma=31~\rm{nm}$, much smaller than the wavelength $\lambda=7 \rm{nm}$. Fig.(\ref{fig:s5}) shows the $|S|$ as a function of $N$ , obtained by sampling the positions of the atoms according to the Gauss distribution with $\sigma=31~\rm{nm}$. $|S|$ is approximately $0.9$ for $N>4$.

As a comparison, we also show the value of $|S|$ for the atomic temperature of $30~\mu\rm{K}$. This temperature is obtained after the polarization gradient cooling. At this temperature, the atomic spatial spread is dominated by thermal fluctuation.

\section{Dark mode purity and linewidth}

\begin{figure}[h]
\includegraphics[width=12cm,angle=0]{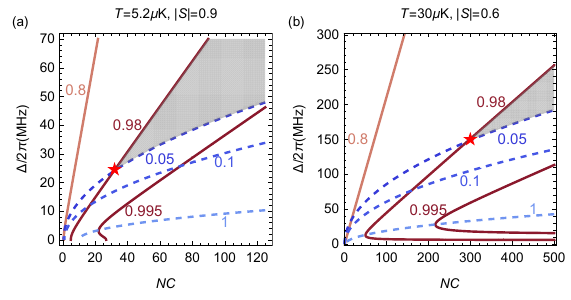}
\caption{\label{fig:s6} Contour plots of the dark mode purity $\mathcal{D}$ (red solid lines) and the relative broadening $\delta\kappa/\kappa$ (blue dashed lines) with the collective cooperativity $NC$ and the atomic detuning $\Delta$ for (a) $|S|=0.9$ and (b) $|S|=0.6$, corresponding to the atomic temperature of $5.2~\mu\rm{K}$ and $30~\mu\rm{K}$, respectively. }
\end{figure}

According to Eq.(\ref{eqs:eom3}), the steady states of the bright and dark modes are written as
\begin{subequations}\label{eqs:ssc}
    \begin{flalign}
        \langle\hat{c}_1\rangle&=\frac{i\frac{S}{\sqrt{2}|S|}\sqrt{\kappa_{\text{in}}}\mathcal{E}_{\text{in}}}{i\delta-\frac{\kappa}{2}+\frac{Ng^2\left(1+|S|\right)}{i\Delta-\gamma/2}}
        =\frac{i\frac{S}{\sqrt{2}|S|}\sqrt{\kappa_{\text{in}}}\mathcal{E}_{\text{in}}}{i(\delta-\delta\omega_1)-\left(\frac{\kappa}{2}+\frac{\kappa}{2}\frac{NC(1+|S|)}{4\Delta^2+\gamma^2}\right)},\\
        \langle\hat{c}_2\rangle&=\frac{i\frac{S}{\sqrt{2}|S|}\sqrt{\kappa_{\text{in}}}\mathcal{E}_{\text{in}}}{i\delta-\frac{\kappa}{2}+\frac{Ng^2\left(1-|S|\right)}{i\Delta-\gamma/2}}=\frac{i\frac{S}{\sqrt{2}|S|}\sqrt{\kappa_{\text{in}}}\mathcal{E}_{\text{in}}}{i(\delta-\delta\omega_2)-\left(\frac{\kappa}{2}+\frac{\kappa}{2}\frac{NC(1-|S|)}{4\Delta^2+\gamma^2}\right)}.
    \end{flalign}
\end{subequations}
The dark mode purity is defined as $\mathcal{D}=n_2/(n_1+n_2)$, where $n_1=|\langle\hat{c}_1\rangle|^2,n_2=|\langle\hat{c}_2\rangle|^2$. When driving the cavity on the dark mode resonance with $\delta=\delta\omega_2$ (Eq.(\ref{eqs:shifts})) substituting into Eq.(\ref{eqs:ssc}), we can plot $\mathcal{D}$ as the collective cooperativity $NC$ and the atomic detuning $\Delta$ in Fig.(\ref{fig:s6}).

From Eq.(\ref{eqs:ssc}b) we can obtain the relative linewidth broadening $\delta\kappa/\kappa$ of the $\hat{c}_2$ mode
\begin{equation}
    \begin{aligned}
         \frac{\delta\kappa}{\kappa}= \frac{NC(1-|S|)\gamma^2}{4\Delta^2+\gamma^2}.
    \end{aligned}
\end{equation}
Ideally the dark mode should have $\delta \kappa/\kappa =0$ since $|S|=1$. However, in practice it always has a small but finite linewidth broadening that arises from the atomic scattering loss, because $|S|<1$ due to the finite spatial extent of atoms, as discussed in the previous section.


Fig.(\ref{fig:s6}) shows the contours of $\mathcal{D}$ (solid lines) and $\delta\kappa/\kappa$ (dashed lines) with $NC$ and $\Delta$. Here we compare the results for $|S|=0.9$ and $|S|=0.6$. The two values of $|S|$ correspond to the atomic temperature of $5.2~\mu\rm{K}$ and $30~\mu\rm{K}$, respectively, obtained with and without Raman sideband cooling. For $|S|=0.9$, we can realize a dark mode with the purity $\mathcal{D}>0.98$ and atomic broadening $\delta\kappa/\kappa< 0.05$ with $NC> 32$, indicated by the red star in Fig. \ref{fig:s6}(a). For a smaller value of $|S|=0.6$, to achieve the same dark mode quality, the requirement becomes $NC>295 $ as shown in Fig. \ref{fig:s6}(b). This demonstrates that a better-structured array, prepared at a lower temperature through the Raman sideband cooling, relaxes the cooperativity requirements needed to achieve a high-quality dark mode.

\section{Photon conversion efficiency}
We obtain the steady states of $\hat{a}_{\pm}$ by solving Eq.(\ref{eqs:eom2})
\begin{subequations}\label{eqs:ssa}
    \begin{flalign}
        \langle\hat{a}_{+}\rangle&=i\sqrt{\kappa_{\text{in}}}\mathcal{E}_{\text{in}}\frac{i\delta-\frac{\kappa}{2}+\frac{Ng^2}{i\Delta-\gamma/2}}{\left(i\delta-\frac{\kappa}{2}+\frac{Ng^2}{i\Delta-\gamma/2}\right)^2-\left(\frac{Ng^2|S|}{i\Delta-\gamma/2}\right)^2},\\
        \langle\hat{a}_{-}\rangle&=i\sqrt{\kappa_{\text{in}}}\mathcal{E}_{\text{in}}\frac{-\frac{SNg^2}{i\Delta-\gamma/2}}{\left(i\delta-\frac{\kappa}{2}+\frac{Ng^2}{i\Delta-\gamma/2}\right)^2-\left(\frac{Ng^2|S|}{i\Delta-\gamma/2}\right)^2}.
    \end{flalign}
\end{subequations}
Define the photon conversion efficiency $\chi$ as the ratio of the output photon number $n_{-}$ of the $\hat{a}_-$ mode over the input photon number $n_{\text{in}}$ in the forward direction
\begin{equation}\label{eqs:chi}
\chi = \frac{n_-}{n_{\rm{in}}}= \frac{\kappa_{\rm{out}} |\langle\hat{a}_{-}\rangle|^2}{|\mathcal{E}_{\text{in}}|^2},
\end{equation}
where $\kappa_{\text{out}}$ is the output mirror coupling rate. Substituting Eq.(\ref{eqs:ssa}b) into Eq.(\ref{eqs:chi}) and setting the cavity on the dark mode resonance with $\delta=\delta\omega_2$, we get
\begin{equation}\label{eqs:conversion1}
    \begin{aligned}
         \chi=\eta_{\text{in}}\eta_{\text{out}}\frac{4|S|^2NC}{(1+\delta\kappa/\kappa)^2\left[4|S|^2NC+(1+3|S|)\delta\kappa/\kappa+(1-|S|)/(\delta\kappa/\kappa)+2(1+|S|)\right]},
    \end{aligned}
\end{equation}
where the input and output mirror coupling efficiency $\eta_{\text{in}}=\kappa_{\text{in}}/\kappa,\eta_{\text{out}}=\kappa_{\text{out}}/\kappa$.

Eq.(\ref{eqs:conversion1}) reduces to a very simple form when $NC \gg 1$ as realized in our experiment, and is given by
\begin{equation}\label{eqs:conversion2}
    \begin{aligned}
         \chi=\eta_{\text{in}}\eta_{\text{out}}\frac{1}{(1+\delta\kappa/\kappa)^2}
    \end{aligned}
\end{equation}

\section{Phase shift governed by the atom array displacement}

In the main text, we analyze the phase shift on the cavity output photon based on the symmetry argument. From Eq.(\ref{eqs:ssa}) we can also explicitly see that the cavity $\hat{a}_{-}$ mode is proportional to $S$. When the atom array is displaced by a distance $X$, the structure factor $S\rightarrow Se^{2 i k X}$ and therefore the phase of the cavity $\hat{a}_{-}$ mode changes by $2k X$. The cavity $\hat{a}_{+}$ mode is determined by $|S|$, hence its phase is unchanged.
\end{document}